\documentclass[twocolumn,showpacs,preprintnumbers,amsmath,amssymb]{revtex4}

\usepackage{amssymb}
\usepackage{amsmath}
\usepackage[dvips]{graphicx}
\usepackage{longtable}
\usepackage{verbatim}
\usepackage{amsfonts}

\newcommand{\C}{\mathbb{C}}
\newcommand{\Z}{\mathbb{Z}}

\newcommand{\h}{\mathcal{H}}
\newcommand{\ten}{{\bf 10}}
\newcommand{\bfive}{{\bf \bar 5}}
\newcommand{\five}{{\bf 5}}
\newcommand{\ufn}{U(1)_{\rm FN}}

\begin{document}

{\hspace*{13cm}\vbox{\hbox{WIS/25/05-Dec-DPP}
  \hbox{hep-th/0512211}}}

\vspace*{-10mm}

\title{Solving Flavor Puzzles with Quiver Gauge Theories}
\author{Yaron~E.~Antebi, Yosef~Nir and Tomer~Volansky}
\address{Department of Particle Physics, Weizmann Institute of Science, Rehovot 76100, Israel}
\date{\today} 

\vspace{2cm}
\begin{abstract}
We consider a large class of models where the $SU(5)$ gauge symmetry and
a Froggatt-Nielsen (FN) Abelian flavor symmetry arise from a $U(5)\times
U(5)$ quiver gauge theory. An intriguing feature of these models is a
relation between the gauge representation and the horizontal charge,
leading to a restricted set of possible FN charges. Requiring
that quark masses are hierarchical, the lepton flavor structure is uniquely
determined. In particular, neutrino mass anarchy is predicted. 
\end{abstract}

\maketitle

\section{Introduction}
\label{sec:introduction}
The charged fermion flavor parameters -- quark masses and mixing
angles and charged lepton masses -- exhibit a structure that is not
explained within the Standard Model (SM). The two puzzling features --
smallness and hierarchy -- are very suggestive that an approximate
horizontal symmetry is at work. The simplest framework that employs
such a mechanism to explain the flavor puzzle is that of the
Froggatt-Nielsen (FN) mechanism \cite{Froggatt:1978nt}. The various
generations carry different charges under an Abelian symmetry. The
symmetry is spontaneously broken, and the breaking is communicated to
the SM fermions via heavy fermions in vector-like
representations. The ratio between the scale of spontaneous symmetry
breaking and the mass scale of the vector-like fermions provides a
small symmetry-breaking parameter. Yukawa couplings that break the FN
symmetry are suppressed by powers of the breaking parameter, depending
on their FN charge.

Model building within the FN framework usually proceeds as follows. 
One chooses a value for the small symmetry-breaking parameter(s), and
a set of FN charges for the fermion and Higgs fields. These choices
determine the parametric suppression of masses and mixing angles. One
then checks that the experimental data can be fitted with a reasonable
choice of order-one coefficients for the various Yukawa couplings.  
Thus all FN predictions are subject to inherent limitations:
\begin{itemize}
\item The FN charges are not dictated by the theory. 
\item The value of the small parameter is not predicted.
\item There is no information on the ${\cal O}(1)$ coefficients.
\end{itemize}
The predictive power of the FN framework is thus limited. There is one
relation among the quark flavor parameters that is independent of the
choice of horizontal charges \cite{Leurer:1992wg}, and
there are three in the lepton sector \cite{Grossman:1995hk}. The
resulting predictions, that suffer from order one uncertainties, are
consistent with the data. Additional relations apply in the
supersymmetric extension of the SM, but to provide new tests of the FN
mechasnim, supersymmetric contributions to flavor changing processes
must be explored, and the universal effects of RGE running should be
small \cite{Nir:2002ah}. The predictive power is sharply enhanced in
the framework of GUT. With an $SU(5)$ gauge symmetry, the number of
independent fermion charges is reduced from fifteen to six. 

To make further progress, one would like to embed the FN mechanism in
a framework where some or all of the inherent limitations described
above are lifted. This may happen in string theory. While realistic
constructions of the supersymmetric SM in string theory are still
under study 
\cite{Ibanez:2001nd,Kokorelis:2002zz,Blumenhagen:2001te,Cascales:2003wn,Blumenhagen:2005mu,Braun:2005ux,Bouchard:2005ag,Giedt:2005vx,Braun:2006ae,Bouchard:2006dn},
much progress has been made in the search for string-inspired
phenomenologically viable extensions of the SM such as the FN
framework. The basic idea is that the FN symmetry is a
pseudo-anomalous $U(1)$ symmetry 
\cite{Ibanez:1994ig,Jain:1994hd,Binetruy:1994ru,Dudas:1995yu,Nir:1995bu}.
Then the small parameter depends on the FN charges and, furthermore, 
if one assumes gauge coupling unification, there is a single
constraint on the FN charges  
that can be translated into a relation between the fermion masses and
the $\mu$-term. This idea is based on ingredients of the heterotic
string and has led to a detailed investigation of the resulting
phenomenology (see, for example, refs.
\cite{Elwood:1997pc,Binetruy:1996xk,Dreiner:2003yr,Harnik:2004yp}.)

On the other hand, we are not aware of any attempt to-date to
construct FN models which arise from D-brane configurations \cite{Cremades:2003qj}.  In
this paper, we take a step in this direction and consider FN models
from quiver gauge theories. These theories arise at low energy as the
effective theories on D-branes placed at singular geometries (see
\cite{Douglas:1996sw,Johnson:1996py,Lawrence:1998ja,Aldazabal:2000sa,He:2004rn,Kiritsis:2003mc}
and references therein). As opposed to the heterotic case, these
theories typically have numerous anomalous $U(1)$'s. The anomalies are
cancelled through the generalized Green-Schwartz (GS) mechanism
\cite{Green:1984sg,Dine:1987xk,Ibanez:1998qp}.  We employ these
anomalous $U(1)$'s as flavor symmetries and construct FN models.

As we show below, the structure of these theories tightly constrains
model building and hence the realization of the FN mechanism.  As a
consequence, we will see that much can be said about the lepton
sector. In particular, within the framework of $SU(5)$ GUT model
with a single FN-symmetry breaking field, there is essentialy a single
viable model which predicts mass anarchy in the neutrino sector.

\begin{figure}[bt]
  \centering
  \includegraphics[scale=0.4]{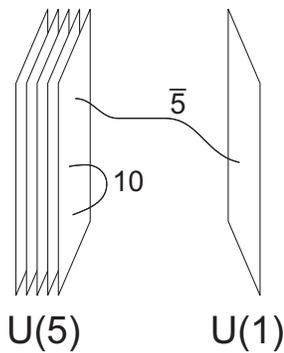}
  \caption{A D-brane construction with an $SU(5)$ gauge group and a
    distinct $U(1)_{\rm FN}$. The fundamental fields are strings
    stretching between
    the two stacks and are thus charged under the FN group, while the
    antisymmetric fields connect only to the $U(5)$ stack and have no
    $U(1)_{\rm FN}$ charge. }
  \label{fig:branes}
\end{figure}

It is worth noting that within the $SU(5)$ GUT model, it is a priori
difficult to get the correct flavor hierarchy altogether.  The reason
for this is that $\ten$ fields arise from open strings with both ends
residing on the same set of $U(5)$ stack of branes, while $\bfive$
fields come from strings with one end on the $U(5)$ stack and the
other on a different brane which may provide the necessary $U(1)_{\rm
  FN}$ symmetry.  This situation is depicted in Fig.
\ref{fig:branes}.  In particular this means that the $\ten$ fields are
not charged under the FN symmetry and there is no hierarchy in the up
sector. As we show below, one can overcome this problem by extending
the gauge symmetry, and the obtained structure is sufficiently
restrictive to provide prediction regarding the neutrino sector.


The paper is organized as follows.  In section
\ref{sec:quiv-gauge-theor} we discuss quiver gauge theories and their
orientifold generalizations.  We further discuss Higgsing and the role
of anomalous $U(1)$'s (and the problems that accompany them) in such
theories.  In section \ref{sec:frogg-niels-models} we investigate the
embedding of the FN mechanism within quiver gauge theories. We argue
that it is difficult to construct models with non-renormalizable
superpotential terms. We focus on the case of a single $U(1)_{\rm
  FN}$ symmetry that arises from anomalous $U(1)$'s. We show that
there are severe restrictions on the possible FN charges, which lead
to a generic constraint on the maximal hierarchy in this framework. In
section \ref{sec:su5-gut-models} we construct FN models in $SU(5)$-GUT
theories with a single FN field. We show that there are only three
possible FN charges for the $\ten$-plets, and two for the
$\bfive$-plets. This situation makes the theory highly predictive. In
particular, requiring that quark masses are hierarchical, a single set
of charges is singled out, leading to a unique flavor structure. The
flavor structure of the lepton sector is fixed, and neutrino mass
anarchy is predicted. We conclude in section \ref{sec:summary}. A
full, consistent, quiver realization for the SU(5) GUT theory, with
emphasis on anomaly cancellation, is presented in Appendix
\ref{sec:an-su5-gut}. We collect the six Tables of our paper at the
very end.
Tables I-III show possible FN charges and
suppression factors in generic quiver theories, while tables IV-VI
give all possible models within our $SU(5)\times SU(5)$ framework. 

\section{Quiver Gauge Theories}
\label{sec:quiv-gauge-theor}
A wide class of ${\cal N}=1$ supersymmetric vacua is obtained by
placing D-branes on singular manifolds of type II strings such as
orbifolds and orientifolds
\cite{Dixon:1985jw,Berkooz:1996dw,Douglas:1996sw,Johnson:1996py,Lawrence:1998ja,Kachru:1998ys,Aldazabal:2000sa,Benvenuti:2004dy,Feng:2000mi,Cvetic:2001nr,Klebanov:1998hh,Martelli:2004wu}.
Placing D-branes at such singularities produces at low energy
conformal gauge theories \cite{Lawrence:1998ja}, while adding
fractional D-branes breaks the conformal symmetry, rendering
a four dimensional chiral gauge theory.

A quiver diagram is an efficient way for describing the gauge theory
obtained from the open string sector (for a review, see
\cite{He:2004rn}). The degrees of freedom of oriented strings can be
described as strings starting and ending on D-branes. Consequently,
the fields in the theory transform in the fundamental of a $U(N_i)$
factor of the gauge group and in the antifundamental of another
$U(N_j)$ factor. It is therefore possible to describe the field theory
by a quiver diagram, where we denote each $U(N_i)$ factor by a node in
the graph and the fields are represented by directed lines connecting
two such nodes. The orientation of the line represents the orientation
of the string: a line coming out of a node corresponding to a $U(N_i)$
gauge group factor stands for a field in the fundamental $\bf{N_i}$,
while a line going into a node corresponding to $U(N_j)$ represents a
field transforming in the antifundamental $\overline{{\bf N_j}}$. A
line originating and ending on the same node, describes a field in the
adjoint representation of the corresponding $U(N)$ factor.

Gauge invariant field combinations, which may be present in the
superpotential, can also be seen using the diagrammatic description. A
field transforming in the fundamental of a given $U(N)$, must interact
with a field in the antifundamental of the same $U(N)$ in order to get
an invariant interaction. In other words, if there is a field coming
into a given vertex, we must also have a field going out of that same
vertex. This has to be the case for all the vertices, so an invariant
interaction is described by a closed loop in the quiver diagram. In
particular, a renormalizable cubic term in the superpotential is
represented in the quiver by a closed triangle.  In general, however,
not every closed loop in the quiver which originates from a certain
geometry appears in the superpotential. The general problem of
extracting the spectrum and superpotential from a given geometry is
still not solved, and only specific examples are known.

\subsection{Orientifold Quivers}
\label{sec:orientifold-models}
Strictly speaking, gauge theories arising from unoriented string
theory are not quivers since the low energy field theory cannot be
described by a directed graph. Nevertheless, these theories may also
be described diagrammatically by `extended' quivers, where the lines
representing the strings are no longer directed. Instead, the two ends
of each string can independently be in either the fundamental or the
antifundamental. Thus each line must be drawn with an arrow at each of
the two ends, indicating what is the representation of the
corresponding string under each of the two gauge group factors.
Unoriented strings with both ends coming out of the same set of branes
may reside in either the symmetric or the antisymmetric combination of
${\bf N}\times \bf{N}$.  Which of these two options is realized is
directly related to the orientifold projection in the original theory:
it is the antisymmetric (symmetric) part for the $SO(N)$ ($Sp(N)$)
orientifold projection. Having said that, we stress that the effective
field theories of unoriented strings on singular manifolds are only
known for $\Z_n$ orbifolds.  Nevertheless, it is expected that the
unoriented nature of the string should lead to the same `extended'
quiver type diagrams in more general singularities.

Again, invariant interactions follow from the (extended) quiver.
It corresponds to a set of lines for which each node has the number
of ingoing fields equal to the number of outgoing fields. 

\subsection{Higgsing}
\label{sec:higgsing}
Higgsing, that is the spontaneous breaking of the gauge group via
vacuum expectation values (VEVs) of scalar fields, changes in general the low
energy description of a given theory. In particular, by Higgsing a
quiver gauge theory, one obtains a (not necessarily supersymmetric) new quiver gauge theory where the
low energy degrees of freedom are bi-fundamentals of the conserved
gauge groups. We can thus write an effective quiver by identifying or
splitting the vertices and lines so that each new node represents an
unbroken $U(N)$ gauge group. 

\begin{figure}[htbp]
  \centering
  \includegraphics[scale=0.4]{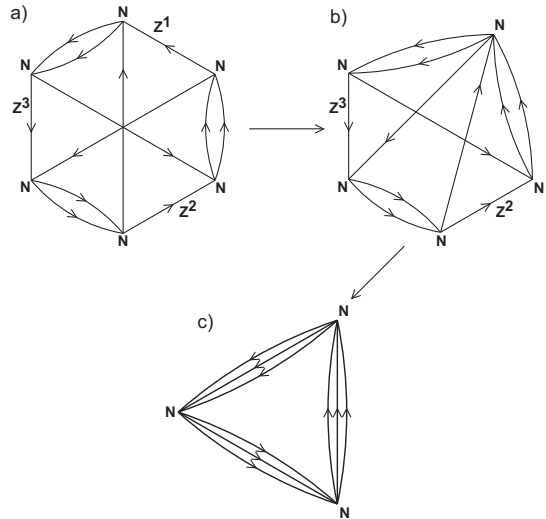}
  \caption{Successive Higgsing of the $Y^{3,0}$ quiver diagram to a
  $\C^3$ quiver: a) The $Y^{3,0}$ quiver diagram; b) The effective
  quiver diagram induced by Higgsing $Z^1\propto\bf1$; c) The
  $\C^3/\Z_3$ quiver induced by 
  Higgsing also $Z^2,Z^3\propto\bf1$.}
  \label{fig:Y30}
\end{figure}

As an example, consider string theory on a $\Z_3$ orbifold of the
conifold (also known as $Y^{3,0}$), with $N$ branes at each
representation of $\Z_3$. The resulting $[U(N)]^6$ gauge theory is
described by the quiver of Fig.~\ref{fig:Y30}a. If we Higgs the theory 
by giving $Z^1$ a VEV proportional to the unit matrix,
\begin{equation}
  \label{eq:11}
  Z^1 = \left( \begin{array}{cccc}
      a & & &  \\
      & a &&  \\
       && a &  \\
      & & & \ddots
    \end{array} \right),
\end{equation}
the corresponding $U(N)\times U(N)$ group is broken to the diagonal
$U(N)$, and the $[U(N)]^5$ gauge theory of Fig.~\ref{fig:Y30}b is
obtained. The $Z^1$ field itself is eaten up by the longitudinal modes
of the broken gauge fields. Higgsing $Z^2$ and $Z^3$ in the same way,
one finds the $[U(N)]^3$ theory described by Fig.~\ref{fig:Y30}c,
which is exactly the quiver of the orbifold $\C^3/\Z_3$
\cite{Benvenuti:2004dy}. 

\begin{figure}[htbp]
  \centering
  \includegraphics[scale=0.4]{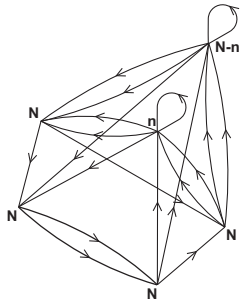}
  \caption{Breaking the $Y^{3,0}$ quiver by a diagonal VEV with two
    different eigenvalues with multiplicities $n$ and $N-n$.}
  \label{fig:Y30other}
\end{figure}

If, instead, we break the $Y^{3,0}$ quiver by choosing a different
form of the VEV for $Z^1$,
\begin{equation}
  \label{eq:12}
   Z^1 = \left( \begin{array}{cccc}
       a &  &  \\
        &  \ddots & \\
      && b &  \\
      &&  & \ddots
    \end{array} \right),
\end{equation}
the symmetry breaking is $U(N)\times U(N) \rightarrow U(n)\times
U(N-n)$. This breaking is no longer supersymmetric and $Z^1$ in not
eaten up completely.  The lines and the nodes split and the quiver
takes a very different form, as can be seen in
Fig.~\ref{fig:Y30other}. Other breaking patterns of $U(N)\times U(M)$
symmetries with bi-fundamentals can be obtained. The most general breaking is
\begin{eqnarray}
  \label{eq:13}
  &U(N)\times U(M)\rightarrow U(N-M+n_0) \times \prod_{i=0}^k U(n_i)&
\end{eqnarray}
with $\sum_{i=0}^k n_i = M$. It leads to more complicated quiver
diagrams which, in general, will not be supersymmetric due to a
non-vanishing D-term.  However, by giving VEVs to a pair of fields in a vector
representation, supersymmetry may easily be preserved.

Different patterns of breaking are generated when the Higgsed field is
in the adjoint representation or, for `extended' quiver, in the
fundamentals of both gauge groups or in the symmetric (or
anti-symmetric) representation. In all cases, the resulting
theories can be described by a new quiver diagram.

\subsection{Anomalous $U(1)$'s}
\label{sec:anomalous-u1s}
Typically, many of the $U(1)$ factors associated with the $U(N)$ gauge
groups are anomalous, with
anomalies cancelled by the generalized Green-Schwartz (GS) mechanism
\cite{Douglas:1996sw,Ibanez:1998qp,Green:1984sg,Dine:1987xk}. In contrast to the case of the
heterotic string, in type II string theory there can be several 
such anomalous $U(1)$ factors. Furthermore, the corresponding gauge
fields are massive independent of whether the symmetry is
spontaneously broken  (the spontaneous breaking is induced by
non-vanishing Fayet-Illiopolous (FI) terms) 
\cite{Lalak:1999bk,Lykken:1998ec}. In the case that the symmetry is
not broken by scalar VEVs, these $U(1)$ factors remain as global
symmetries \cite{Witten:1984dg} and the corresponding $SU(N)$
factors remain good gauge symmetries. Thus each node in the quiver has
a corresponding (local) global (non-)anomalous $U(1)$.

Phenomenologically, these $U(1)$ factors pose considerable problems to
model building, essentially since the global $U(1)$ symmetries of the
SM or its supersymmetric extensions (SSM) are not directly related to
the local gauge groups. Consider, as an example, a supersymmetric
quiver theory which contains the $SU(2)_W$ gauge group
\cite{Ibanez:2001nd}.  Under the corresponding $U(1)$, all the
doublets ($Q_i$, $L_i$, $H_u$ and $H_d$) are charged $\pm1$. The
superpotential of the SSM has the following form:
\begin{equation}
  \label{eq:17}
  W = Y_{ij}^dH_dQ_id_j + Y_{ij}^uH_uQ_iu_j + Y_{ij}^LH_dL_ie_j +\mu
  H_uH_d. 
\end{equation}
It is clear that, to allow for $Y^u\neq0$ and $Y^d\neq0$, the $U(1)$
charges of $H_u$ and $H_d$ must be the same. But then the $\mu$ term
is forbidden. A similar problem exists in the $SU(5)$ GUT models
\cite{Blumenhagen:2005mu, Blumenhagen:2001te} and their extensions, as
we discuss in section \ref{sec:su5-gut-models}.

There are three possible solutions to the above problem:
\begin{enumerate}
\item The particle content of the low energy theory is extended in
  such a way that the symmetry is realized.  For example, an extended
  higgs sector \cite{Ibanez:2001nd} enlarges the global symmetry. 
\item The $U(1)$ is broken spontaneously.  The only way to do this
  without breaking the associated $SU(N)$ gauge group is by letting a
  singlet composed of $N$ fundamentals to obtain a VEV.  This breaks
  the $U(1)$ down to a $\Z_N$.  Such composite singlets may exist if
  they are charged under a different, confining gauge group.
\item The anomalous $U(1)$ is broken by non perturbative effects.
  Depending on the matter content, such effects break the symmetry
  down to $\Z_{kN}$ for some integer $k$.  
\end{enumerate}
While difficult to analyze, these solutions allow realistic extensions
of the SSM through quivers. We discuss these possibilities further in
section \ref{sec:su5-gut-models}.

\section{The FN Mechanism from Quivers}
\label{sec:frogg-niels-models}
\subsection{Renormalizable or Non-Renormalizable}
\label{sec:renor}
The Froggatt-Nielsen (FN) mechanism \cite{Froggatt:1978nt} provides an
explanation to the flavor puzzle using a horizontal symmetry. Quarks 
and leptons of various generations are charged differently under a
symmetry $\cal H$ which is spontaneously broken. The simplest
realization is through the use of a single Abelian $U(1)$ which is
broken by the VEV of a scalar field $S$. To allow for ${\cal
  H}$-invariant interaction terms involving the SM Higgs and fermion
fields, powers of $S$ must of be involved, depending on the ${\cal
  H}$-charge of the Yukawa interaction. Thus, non-renormalizable
interaction  terms arise, suppressed by inverse powers of  $M_V$, the
scale at which the breaking of ${\cal H}$ is communicated to the
SM. The effective Yukawa interactions are then suppressed by powers of
$\langle S\rangle/M_V<1$, and are characterized by smallness and
hierarchy, as required phenomenologically.

One way of embedding the FN mechanism in string theory would be by
identifying the scale $M_V$ with the string scale, and obtaining a
superpotential that has most of its terms non-renormalizable (the top,
and perhaps other third generation Yukawa couplings, being the
exception). This goal is difficult to achieve. Quiver gauge theories
which arise from D-branes at singularities have a superpotential
determined fully by the geometry. Geometrically engineering the
required superpotential is hard. Most of the geometries which are
under control arise from orbifold singularities \cite{Douglas:1996sw}
or toric geometries
\cite{Benvenuti:2004dy,Feng:2000mi,Martelli:2004wu}. (These two
classes are, of course, not separate: Abelian orbifolds are toric.) In
the first type of geometry, the superpotential is obtained by
truncating ${\cal N}=4$ superconformal Yang-Mills which has cubic
interactions.  For the second type, in the case of $Y^{p,q}$ geometries,
one obtains also quartic interactions. In any case, it is difficult to
construct a quiver theory with a superpotential that has most of its
terms non-renormalizable.

A second approach is to construct a renormalizable model above the
scale $M_V$ which, at low energy, produces the required
interactions. Further motivation for such a construction comes from
the possible identification of the FN symmetry with an anomalous
$U(1)$, as discussed below. 

\begin{figure}[t]
\centering
\includegraphics[scale=0.7]{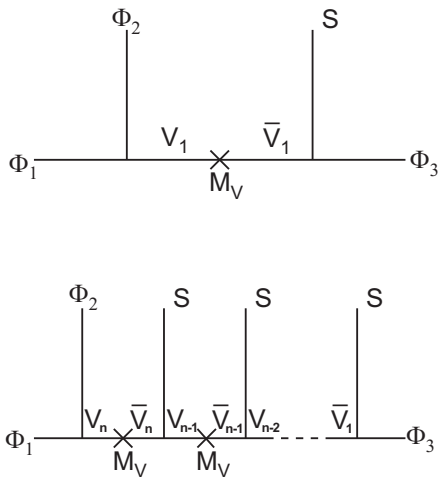}
\caption{Diagrams for generating interactions suppressed by
  different factors $<S>/M_V$ using only cubic interactions.}
\label{fig1}
\end{figure}

A renormalizable model is easy to construct with the introduction of
additional vector-like massive fields. As an example, consider a
non-renormalizable superpotential term of the form
\begin{equation}
  \label{eq:14}
  W = \left(\frac{S}{M_V}\right)^n \Phi_1\Phi_2\Phi_3.
\end{equation}
To generate such interaction, we introduce additional vector-like
massive fields $V_{k},\overline{V}_{k}$ $(k=1,..,n)$, with masses at
the scale $M_V$ and the following charges under the horizontal
symmetry: 
\begin{eqnarray}
  \label{eq:15}
  &\h(\Phi_1)=n,\ \h(\Phi_2)=0,\ \h(\Phi_3)=0,&\nonumber\\
  &\h(S)=-1,\ \h(V_k)= -\h(\overline{V}_k) = -k.&
\end{eqnarray}
Taking the renormalizable superpotential to be
\begin{equation}
  \label{eq:1}
  W = \Phi_1\Phi_2V_{n} + \overline{V}_nV_{n-1} S + ... + \overline{V}_1 S \Phi_3 +
  \sum_i M_V\overline{V}_iV_i, 
\end{equation}
and integrating out $V_i$ and $\overline{V}_i$, one finds the required
interaction, eq. \eqref{eq:14}. Fig.~\ref{fig1} shows the relevant
diagrams which generate this low energy interaction for $n=1$
and for general $n$.

\subsection{Identifying the FN Symmetry}
\label{sec:fnquiver}
To relate the FN mechanism to a quiver field theory, one needs to
identify a global symmetry that can play the role of a viable
horizontal symmetry. We limit our search of viable models to theories
with the following features:
\begin{itemize}
\item The FN symmetry arises from anomalous gauged $U(1)$'s in the
  quiver. (In particular, we do not consider global symmetries of the
  open string sector that are related to isometries in the
  dual gravitational theory.)
\item The spontaneous breaking of the symmetry comes from a VEV of a
  single field, $S$.
\item The ratio of the two relevant energy scales is small,
\begin{equation}
\epsilon\equiv\langle S\rangle/ M_V \ll 1,
\end{equation}
allowing for smallness and hierarchy in the effective Yukawa couplings.
\end{itemize}

We note several points that hold generically in a FN mechanism
that arises from a quiver theory using an anomalous $U(1)$:
\begin{enumerate}
\item For $S$ to be charged under an anomalous $U(1)$, it must either
  be stretched between two distinct vertices, or have its two ends on
  the same vertex but with both ends sitting in the same
  representation (either fundamental or antifundamental). The second
  situation is possible only in the orientifold case.  Since $S$ is
  generically charged under the full $U(N)[\times U(M)]$, and not just
  under the anomalous $U(1)$ factors, it necessarily breaks some of
  the gauge groups.  This means that {\it the FN
    mechanism requires an extended group}.
\item The charge of a given field under the FN symmetry is fixed by
  its representation under the corresponding non-Abelian gauge groups.
  For example, assume $S$ resides in the $({\bf N_L}, {\bf \bar N_R})$
  representation of a $U(N_{\rm L})\times U(N_{\rm R})$ gauge group.
  It is neutral under the sum of the two $U(1)$ factors, $U(1)_{\rm
    L+R}$, but has a charge $+2$ (in a specific normalization) under
  the difference, $U(1)_{\rm L-R}$. Thus we must identify $U(1)_{\rm
    FN}$ with $U(1)_{\rm L-R}$. Any field in the $(\bf N_L, 1)$
  representation has then a FN charge of $+1$, while a field in the
  $(\bf \bar N_L, \bf \bar N_R)$ representation is neutral under the
  FN symmetry.
\item There may be other fields in the theory obtaining VEVs that
  break additional gauge groups or $U(1)$'s. Our assumptions above
  mean that those VEVs are of the order of $M_V$ and do not contribute
  to the hierarchy. To distinguish between these fields and $S$, one
  may write the effective quiver after the Higgsing of all fields
  except $S$.
\end{enumerate}

\subsection{FN charges}
\label{sec:fncharges}
As explained above, the charges of the various fields under the FN
symmetry are very restricted.  This, in turn, affects the possible
hierarchical structure within the quiver theory. In fact, the
strongest suppression possible is by $\epsilon^3$
($\epsilon\equiv\langle S\rangle/M_V$) and even this suppression is
unlikely to actually appear, as we explain below.

\begin{figure}[bt]
  \centering
  \includegraphics[scale=0.4]{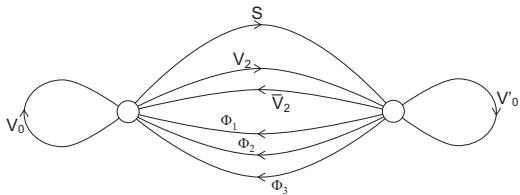}
  \caption{A directed quiver diagram for an $\epsilon^3$-suppression
  of the effective $\Phi_1\Phi_2\Phi_3$ term in the superpotential.}
  \label{fig:S3Orbifold}
\end{figure}

Consider first oriented strings. As explained above, for directed
quivers, $S$ must be stretched between two distinct branes and so its
charge is $(+1,-1)$ under the corresponding $U(1)_{\rm L}\times
U(1)_{\rm R}$. The FN symmetry is then $U(1)_{\rm FN} = U(1)_{\rm
  L-R}$ under which $S$ is charged $+2$. Any other field can be
charged with one of the following: $(0,0), (\pm1,0), (0, \pm 1), (\pm
1,\mp 1)$.  Thus the strongest suppression of an effective Yukawa term
of the form $\Phi_1\Phi_2\Phi_3$ is obtained when all three fields are
charged $(-1,+1)$, giving an $\epsilon^3$ suppression. The corresponding
quiver diagram diagram is drawn in Fig.~\ref{fig:S3Orbifold}.

While an $\epsilon^3$-supperssion can, in principle, be generated, it
is unlikely to be relevant in practice, since it requires that all
three fields $\Phi_i$ transform in the same way under the
entire gauge group. In particular, there is no such case in  
$SU(5)$ GUT models, as we discuss further in section
\ref{sec:su5-gut-models}.  

\begin{figure}[t]
  \centering
  \includegraphics[scale=0.4]{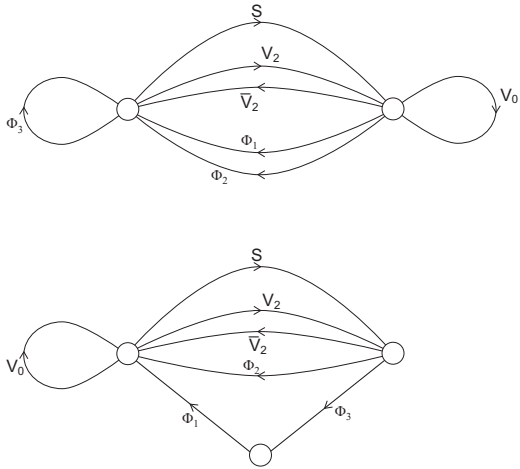}
  \caption{Directed quiver diagrams for an $\epsilon^2$-suppression of
    the effective $\Phi_1\Phi_2\Phi_3$ term in the superpotential.}
  \label{fig:S2Orbifold}
\end{figure}

To get an $\epsilon^2$-suppression, there are two possibilities for the
charges: $(+1,-1), (+1,-1), (0,0)$ or $(+1,-1), (+1,0), (0, -1)$.
Examples for the two sets of charges are drawn in Fig.~\ref{fig:S2Orbifold}.

\begin{figure}[t]
  \centering
  \includegraphics[scale=0.6]{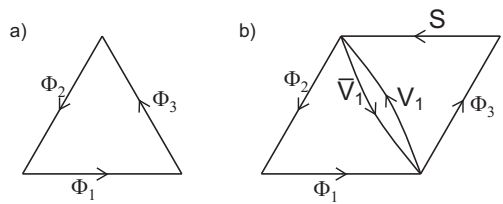}
  \caption{Directed quiver diagrams which lead to
  (a) unsuppressed effective Yukawa coupling, and (b)
  $\epsilon$-suppression of the effective Yukawa coupling.}
\label{fig2}
\end{figure}

There are five sets of charges that yield an
$\epsilon$-suppression. These (and the other sets of charges that
yield suppression) are presented in Table
\ref{tab:charges-orbifolds}. The quiver diagram that corresponds to
the set $(0,0),(-1,0),(0,+1)$ is depicted in Fig.~\ref{fig2}(b).
Figure \ref{fig2}(a) shows the simplest triangle which produces a
renormalizable (that is unsuppressed) Yukawa coupling.

\begin{figure}[htbp]
  \centering
  \includegraphics[scale=0.4]{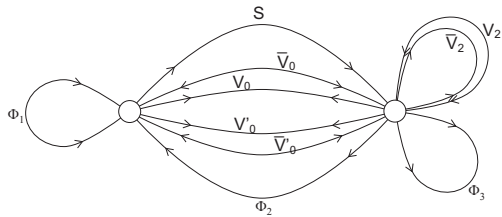}
  \caption{A quiver diagram in the orientifold case that gives an
  $\epsilon^3$-suppression.} 
  \label{fig:S3orient}
\end{figure}

For orientifolds, in addition to the above $U(1)_{\rm L}\times
U(1)_{\rm R}$ charges, there could be fields with charges $(\pm2,0)$,
$(0,\pm2)$ and $(\pm1,\pm1)$. Furthermore, $S$ itself can have charges
$(+1,-1)$ (similar to the oriented string), $(+1,+1)$ or $(\pm2,0)$.
The breaking patterns of the gauge groups will be different for these
three choices. The $U(1)_{\rm FN}$ is then $U(1)_{\rm L-R}$,
$U(1)_{\rm L+R}$ or $U(1)_{\rm L}$, respectively. For the first case,
table \ref{tab:charges-orientifolds} enumerates the possible
suppression factors that arise in addition to the orbifold case of
table \ref{tab:charges-orbifolds}.  Note the additional configuration
with an $\epsilon^3$ suppression, whose relevant quiver diagram is
shown in Fig.~\ref{fig:S3orient}. The second case, $U(1)_{\rm L+R}$,
can be easily obtained from the first one, $U(1)_{\rm L-R}$, by
multiplying the $U(1)_{\rm R}$ charges by minus one. Finally, the
third case, $U(1)_{\rm L}$, again exhibits a configuration of
$\epsilon^3$-suppression, as can be seen in table
\ref{tab:charges-orientifolds3}. Here too, this configuration requires
all three fields to be in the same representation of the non-Abelian
gauge groups.

Our results in this section show the strong predictive power that is
added to the generic Froggatt-Nielsen mechanism when embedded in
string theory. In fact, the constraints are so strong -- {\it i.e.}
the strongest suppression of a Yukawa coupling is third order in a
small parameter and, for practical purposes, probably only second
order -- that one may wonder whether our framework gives rise to any
viable flavor model at all. Indeed, in the next section we show that
to construct viable models, one has to invoke (rather plausible)
non-perturbative effects. These effects relax some of the constraints
that we presented in this section and, in particular, allow an
$\epsilon^4$-suppression of the Yukawa couplings in the case of
$SU(5)$ GUT models.

\section{$SU(5)$ GUT Models and Neutrino Mass Anarchy}
\label{sec:su5-gut-models}
In this section we consider $SU(5)$ GUT models with a FN flavor
symmetry. We search for viable models that arise from quiver gauge
theories. The theory turns out to have intriguing
implications for the neutrino sector. 

By an $SU(5)$ GUT model we mean that there is a range of energy scales
where the gauge group is $SU(5)$, with matter fields that transform as
${\bf5}$, ${\bf\bar 5}$, and ${\bf10}$. The presence of an
antisymmetric multiplet of the gauge group requires that we consider
orientifold theories and choose the appropriate projection. This
projection is the one that results in an $SO(N)$ gauge group(s). 

Our analysis focusses on the energy scale just above the GUT breaking 
scale. In general there can be many fields that break the various
gauge groups that are present in the quiver theory. However, as
discussed in the previous section, we consider a scale that is low
enough so that the only field  to play a relevant role in the breaking
of the FN symmetry and possibly in breaking of a
larger gauge group into $SU(5)$ is the FN field. 

\subsection{General Considerations and Predictions}
The strongest mass hierarchy in the various fermion mass matrices
appears in the up sector. Thus, a minimal requirement that we put on
viable models is that they produce an up mass hierarchy. This
requirement significantly narrows down the possible configurations.

There are two options regarding the $SU(5)$ gauge group. First, it
could be related to a single node, namely it is a subgroup of a single
$U(N)$ symmetry. In this scenario, the fields transforming as $\bf 10$
must have both ends on the $SU(5)$-related node. Consequently they all
carry the same $\ufn$ charge, regardless of whether $\ufn$ is (i) a
subgroup of the same $U(N)$, (ii) unrelated to this $U(N)$ or (iii) a
subgroup of $U(N)\times U(M)$, where $U(M)$ is related to a different
node. Thus, this scenario gives rise to up mass anarchy (that is, no
special structure in the up mass matrix) and is therefore
phenomenologically excluded. 

The second scenario has the $SU(5)$ gauge group related to two nodes,
namely it is a subgroup of a $U(N_L)\times U(N_R)$ symmetry. The FN field
must be in the bifundamental $({\bf N_L},{\bf\bar N_R})$ of the two
nodes. This case has a rich flavor structure and is the only one that
can lead to phenomenologically viable models. 

The simplest models have the following pattern of gauge symmetry
breaking: $SU(5)\times SU(5)\rightarrow SU(5)_{\rm diag}$. We focus on
this class of models. (More complicated breaking patterns
have a similar hierarchical form, but involve extended particle
content.) The $\bfive$-plets then transform under the $SU(5)\times
SU(5)\times U(1)_{\rm L}\times U(1)_{\rm R}$ as either
$(\bfive,1)_{-1,0}$ or $(1,\bfive)_{0,-1}$. The $\ten$-plets transform
as either $(\ten,1)_{+2,0}$ or $(1,\ten)_{0,+2}$ or
$(\five,\five)_{+1,+1}$.  The $H_u(\bf5)$ field transforms as
either $(\five,1)_{+1,0}$ or $(1,\five)_{0,+1}$.

While $S$ breaks the $U(1)_{\rm FN} = U(1)_{\rm L-R}$ symmetry, it
leaves $U(1)_{\rm L+R}$ as a global symmetry in the $SU(5)_{\rm diag}$ 
theory, under which the fields are charged as
\begin{equation}
  \label{eq:16}
  \bfive(-1),\,\,\,\ten(+2),\,\,\,H_d(-1),\,\,\,H_u(+1).
\end{equation}
This symmetry is flavor diagonal, with charges that are determined
solely through the $SU(5)$ representation. As mentioned
in section \ref{sec:anomalous-u1s}, such $U(1)$'s are generic and face
strong phenomenological constraints. Clearly, the symmetry
of eq. (\ref{eq:16}) is not a symmetry of the $SU(5)$ GUT model. In
particular, no up-type masses are allowed with the above symmetry
\cite{Blumenhagen:2005mu,Blumenhagen:2001te}. One can try to overcome
this problem by introducing a composite $H_u$ field of charge $-4$
\cite{Blumenhagen:2001te}. This solution to the up mass(lessness)
problem comes, however, at the cost of two new problems: First, the
$H_uH_d$ term is now forbidden, rendering the higgsinos
massless. Second, the $\bfive\bfive H_uH_u$ terms are forbidden,
rendering the neutrinos massless. 

As discussed in section \ref{sec:anomalous-u1s}, there are three
possible solutions to this problem, which involve either extending the
particle content and the symmetry of the low energy theory, or
breaking the symmetry either spontaneously or
non-perturbatively. Before going into details, we observe that, in
fact, the quiver theory gives interesting predictions that are
independent of which solution to the up mass problem is employed. 

Indeed, since the $\bfive$ fields carry $U(1)_{\rm L}\times
U(1)_{\rm R}$ charges of either $(-1,0)$ or $(0,-1)$, at least two of
them have the same FN charge. Thus, there must be at least
quasi-anarchy (that is two non-hierarchical masses and one mixing
angle of order one) in the neutrino sector. In half of the models 
all three $\bfive$ fields have the same FN charge, leading to complete
neutrino mass anarchy (no hierarchy in the masses and all three angles
of order one). In addition, this situation, where a maximum of two
possible FN charges are available to the three $\bfive$ fields, has
implications for the down sector: either one or all three (in
correspondence to quasi- or full-anarchy in the neutrinos) down mass
ratios are of the same order as the corresponding mixing angles ({\it
  e.g.} $m_s/m_b\sim|V_{cb}|$). 

A word of caution is, however, in order. The above restrictions on the
possible FN charges can be evaded if the $\ten$ and $\bfive$-plets are
composite fields \footnote{We thank Micha Berkooz for drawing our
  attention to this point.}. In such a case, it is possible for all
fields to carry various charges under additional $U(1)$ factors, which
may play the role of the FN symmetry. Such a possibility, however,
complicates the theory considerably and does not seem to be
attractive, especially since the effective theory cannot be described
by a quiver. We thus assume that the SM fermions are elementary
fields, while condensates can either break the $U(1)_{\rm L+R}$
symmetry or generate effective Higgs fields.

We next discuss the possibilities for solving the $U(1)_{\rm L+R}$
problem. We do so in the specific context of $SU(5)$ GUT, but the
solutions can be straightforwardly generalized to other gauge groups.

\subsection{Extending the Higgs Sector}
To allow for up-quark, higgsino and neutrino masses in a model that
has the $U(1)_{\rm L+R}$ symmetry, one needs to add matter fields. The
simplest extension has a second pair of Higgs doublets\cite{Ibanez:2001nd}. The $U(1)_{\rm
  L+R}$ charges of the four Higgs fields are as follows:
\begin{equation}
  \label{eq:18}
  H_u(-4),\,\, H_d(-1),\,\,\,\, h_d(+4),\,\, h_u(+1).
\end{equation}
Here $H_d$ and $h_u$ are fundamental fields, while $H_u$ and $h_d$
are composite. 

One can now distinguish between models according to the FN charges of
the matter fields, that is the three $\ten_i$, the three $\bfive_i$,
the two elementary Higgs fields $H_d$ and $h_u$, and the two composite
Higgs fields $H_u$ and $h_d$. There are 640 different sets of
$U(1)_{\rm FN}=U(1)_{\rm L-R}$ charges. They are listed in table
\ref{tab:2HM-composite}.

We now impose phenomenological requirements to see, first, if there
are viable models and, second, if these models make further
predictions. It turns out that requiring that the quark masses are
hierarchical is enough to select a {\it single}
flavor structure for all fermions and, in particular, predict the
flavor structure of the lepton sector.

We first consider the up sector. We require that the three up-type
quarks have masses and that these masses are hierarchical. This means
that no two $\ten$-plets are allowed to carry the same FN charge. Out
of the ten different sets of charges for the $\ten$-plets $T_i$, only
one fulfills this requirement, that is $T_5$ of table
\ref{tab:2HM-composite}. Furthermore, the up sector couples to the
composite $H_u$. In order that the up masses do not vanish, $H_u$ must
carry charge $(-4,0)$ under $U(1)_{\rm L}\times U(1)_{\rm R}$. Thus,
of the four sets $U_i$, only $U_1$ and $U_2$ are viable charge assignments.

We next consider the down sector. We require that the three down-type
quarks have masses and that these masses are hierarchical. Out of the
four different sets of charges for the $\bfive$-plets $F_i$, only one
fulfills this requirement, that is $F_1$. Furthermore, the down sector
couples to the elementary $H_d$. $H_d$ must be connected to the same
node as the $\bfive$. Thus, of the four sets $D_i$, only $D_1$ and
$D_2$ are viable choices. 

We are therefore left with a unique set of $U(1)_{\rm L-R}$ charges,
up to the choice of the charges for $h_d$ and $h_u$. This freedom,
however, only affects the $\mu$-terms and the overall scale of the
neutrino masses.  The flavor structure is unaffected by this
choice. Taking the configuration ``$T_5 F_1 D_1 U_1$,'' we obtain the
following parametric suppressions for the various entries in the
fermion mass matrices:
\begin{eqnarray}
  \label{eq:19}
  M_u &\sim& \langle H_u\rangle
  \left(\begin{array}{ccc}
      \epsilon^4&\epsilon^3&\epsilon^2\\ \epsilon^3&\epsilon^2&\epsilon \\ 
      \epsilon^2&\epsilon&1  
    \end{array}
  \right),
  \\
  M_d &\sim& \langle H_d\rangle
  \left(\begin{array}{ccc}
       \epsilon^2&\epsilon^2&\epsilon^2 \\ \epsilon&\epsilon&\epsilon
  \\ 1&1&1 
    \end{array}
  \right),
  \\
  M_\nu &\sim& \frac{\langle h_u\rangle^2}{M} 
  \left(\begin{array}{ccc}
      1&1&1 \\ 1&1&1 \\  1&1&1 \\ 
    \end{array}
  \right).
\end{eqnarray}

There are other ways to extend the matter content in order to
incorporate the $U(1)_{L+R}$ as a the symmetry of the theory. One way
to generate the $\mu$-term for the $H_uH_d$ fields, without
introducing another pair of higgs fields, can be achieved in a similar
fashion to the NMSSM.  One assumes $H_u$ is a
condensate $\bfive\bfive\bfive\bfive$ and introduces another field $T$
which is a $\five\five\five\five\five$ condensate of the same strongly
coupled gauge group. Then the coupling $T H_u H_d$ is allowed, while
the coupling $T^3$ which breaks the $U(1)_{\rm L+R}$, may be generated
by non-perturbative effects below the scale at which the $SU(5)_{\rm
  diag}$ becomes strong, or may not be generated at all, as in the
MNSSM
\cite{Panagiotakopoulos:1999ah,Panagiotakopoulos:2000wp,Dedes:2000jp}.
We do not discuss this idea further, and just note that, as before,
the neutrino sector is predicted to be anarchical.

\subsection{Breaking $U(1)_{\rm L+R}\to\Z_5$}
Another way to get the up Yukawa terms is by breaking the symmetry
through non-perturbative effects. In general, operators which violate
an anomalous symmetry are generated non-perturbatively and can be
calculated using the holomorphic structure of the superpotential. It
is important to note that the Lagrangian of the minimal $SU(5)$
posseses a $\Z_5$ symmetry under which the fields are charged as in
eq. \eqref{eq:16}. This $\Z_5$ is a subgroup of $U(1)_{\rm L+R}$. In
our case, of an $SU(5)_{\rm L}\times SU(5)_{\rm R}$ gauge group, it is
simple to see that if in one of the nodes, say the left, there is only
a single antisymmetric $\ten$ (and another $\bfive$ to cancel gauge
anomalies), then the related $\Lambda_{\rm QCD}^b$, where $b$ is the
coefficient in the $\beta$ function, is charged 5 under the anomalous
symmetry. Therefore, instantons, if exist, break this $U(1)$ down to
$\Z_5$ and can generate masses in the up sector.  Unfortunately, at
the field theory level, no non-perturbative terms are generated in the
superpotential even in this case. The reason is that the number of
flavors in this model is $\geq5$\cite{Poppitz:1995fh,Pouliot:1995me}.

Nevertheless, it could be that non-perturbative corrections which
break the anomalous $U(1)$ arise already at the string level. Such
corrections cannot be calculated explicitly. However, they break the
$U(1)$ in a way that follows from the anomaly and hence may generate
the required up quark Yukawa couplings. Assuming that these terms are
indeed generated in this way, we consider the set of models where
there is a single antisymmetric representation on one of the nodes.  The
list of the 80 possible charge assignments is given in table
\ref{tab:instbreak}. 

Just as for the previous case, most of the possible charge assignments 
lead to a phenomenologically excluded models.  We find again that up
mass hierarchy requires that the three $\ten$-plets have three
different charges  ($T_5$), and down mass hierarchy requires that the
$\bfive$ and $H_d$ fields connect to the left node ($F_1D_1$) in which
the $U(1)$ is broken down to $\Z_5$.
Both possible charge assignments for $H_u$ give non-vanishing
hierarchical masses.  Note, however, that $U_2$ gives an overall
suppression of order $\epsilon$ in the up sector.
The neutrino flavor structure is again anarchical (independent of the
choice $U_i$, though the overall scale depends once again on this choice).

The last class of models involves spontaneous breaking of
$U(1)_{L+R}$. This can be achieved by adding a condensate of five
fields in the fundamental representations of one $SU(5)$ factor. This
condensate includes a singlet of the non-Abelian gauge group with a
$U(1)_{\rm L+R}$ charge $+5$, which we denote by $K$. By giving $K$ a
VEV, we break $U(1)_{\rm L+R}\to\Z_5$ and allow up type mass terms. A
second, conjugate, field $\bar K$ is needed in order to allow for a
mass term. The list of the 320 different charge sets is given in table
\ref{tab:spontanbreak}. The majority of the sets of charges are not
viable. Only two models are viable: $T_5F_1D_1U_1K_1$ and
$T_5F_1D_1U_2K_1$, where the latter, as in the previous case, has an
overall suppression in the up sector.  Once again, anarchy is
predicted for the neutrino sector.

\begin{figure}[htbp]
  \centering
  \includegraphics[scale=0.4]{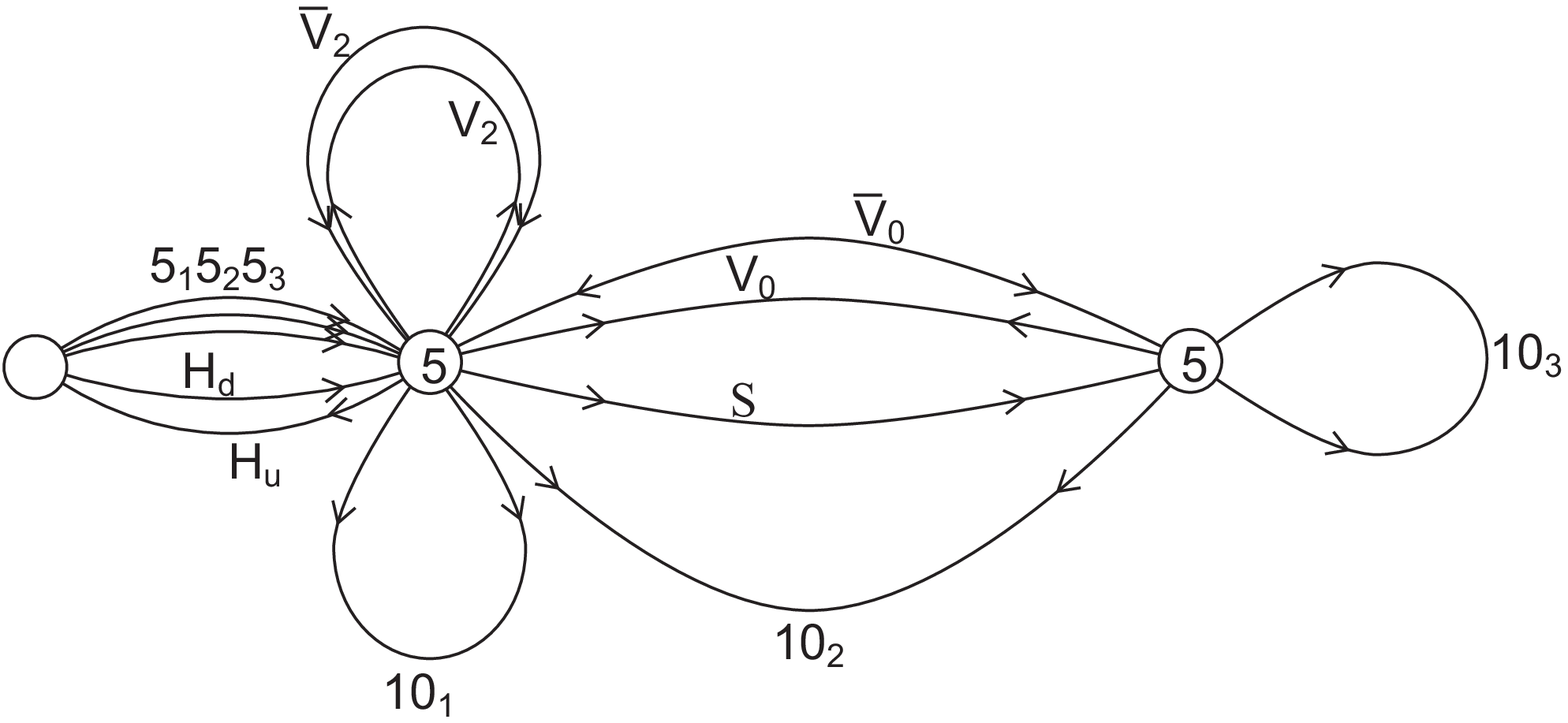}
  \caption{The unique configuration for the SU(5) GUT fields.}
  \label{fig:swhale}
\end{figure}

In all three classes of models, we arrived at an essentially
equivalent configuration for the matter content. The differences
between the three models are just with respect to new fields that are
added to solve the $U(1)_{\rm L+R}$ problem. This unique
configuration, which leads to the flavor structure of eq.
(\ref{eq:19}), is presented in the quiver of
Fig.~\ref{fig:swhale}. (The theory described by this quiver diagram
suffers from non-Abelian gauge anomalies. A non-anomalous extension is
presented in appendix \ref{sec:an-su5-gut} and
Fig.~\ref{fig:ewhale}.) This theory produces, at low energy, the minimal
$SU(5)$ with the correct hierarchy in the up and down sector and with
the predicted neutrino anarchy.  We stress that, given our
assumptions, this theory is unique and thus the anarchy is predicted.

\section{Summary}
\label{sec:summary}
The fermion flavor parameters of the Standard Model have a special,
non-generic structure, that finds no explanation within the Standard
Model. The hierarchy and the smallness of the Yukawa couplings are
suggestive of an approximate symmetry. The Froggatt-Nielsen (FN) mechanism
is a simple and attractive realization of this idea. The mechanism is, 
however, limited in its predictive power. In particular, the FN
charges are not dictated by the theory, and there is only information
on the parametric suppression, but not the order one coefficients, of
the Yukawa couplings. In this work, we investigated whether the
embedding of the FN mechanism in string theory improves its predictive
power.  

Specifically, we examined quiver gauge theories which arise at low
energy from type II string theory with D-branes placed on singular
manifolds. The quiver gauge theories can be described by
(non-)directed graphs for (un)oriented strings, in which the nodes
represent the gauge groups ($U(N)$, $SO(N)$ or $Sp(N)$), while lines
represent matter fields charged under the corresponding gauge groups.

In general, these theories contain several anomalous $U(1)$ symmetries
whose anomaly is canceled by a generalized GS mechanism.  The unbroken
global symmetries can be used to generate the hierarchy of the Yukawa
couplings, thus realizing the FN mechanism.  Since charges of the
matter fields under these $U(1)$ factors are fixed by their
representation under the gauge groups, the FN charges are fixed.
Consequently, one of the inherent limitations of FN models is removed.

Preciesly because it is highly predictive, the above framework does
not easily lend itself to the construction of viable models. We have
discussed this problem and its possible solutions. Concentrating on a
large class of $SU(5)$ GUT models, we have demonstrated the predictive
power of quiver gauge theories. Requiring mass hierarchy in the up
sector, we showed that the $SU(5)$ must come from an extended product
group such as $U(5)\times U(5)$. Furthermore, there must be either
quasi- or full-anarchy in the neutrino sector and either one or all
three down mass ratios are of the same order as the corresponding mixing angles
({\it e.g.} $m_s/m_b\sim|V_{cb}|$).  Further requiring mass hierarchy
in the down sector, the FN charges of all matter fields are
essentially fixed. Consequently, the lepton flavor structure is
predicted and, in particular, there is anarchy (that is, no special
structure) in the neutrino sector.

\section{Acknowledgement}
We are grateful to Micha Berkooz and Yael Shadmi for collaboration in
early stages of this work.  We thank Ofer Aharony, Tom Banks, Michael
Dine, Guy Engelhard, Amihay Hanany, Eric Poppitz, Guy Raz, Martin
Schmaltz and Adam Schwimmer for useful discussions. The work of YN and
TV is supported by a grant from the United States-Israel Binational
Science Foundation (BSF), Jerusalem, Israel. YN~is supported by the
Israel Science Foundation founded by the Israel Academy of Sciences
and Humanities, by G.I.F., the German--Israeli Foundation for
Scientific Research and Development, by EEC RTN contract
HPRN-CT-00292-2002, and by the Minerva Foundation (M\"unchen).

\appendix

\section{An SU(5) GUT Quiver}
\label{sec:an-su5-gut}
\begin{figure}[htbp]
  \centering
  \includegraphics[scale=0.4]{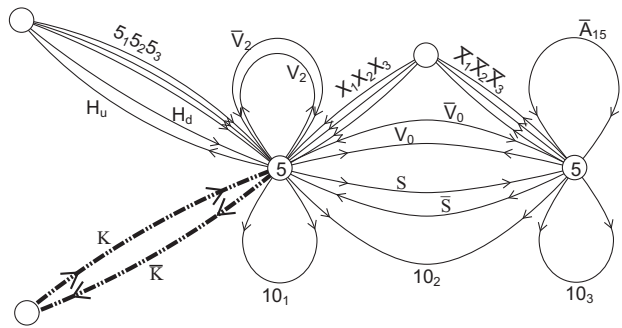}
  \caption{An SU(5) GUT Quiver.}
  \label{fig:ewhale}
\end{figure}

As we showed in section \ref{sec:su5-gut-models}, all three classes of
models have an equivalent configuration for the matter content, which
leads to the flavor structure of eq.  (\ref{eq:19}).  A non-anomalous
quiver which realizes this structure is presented in
Fig.~\ref{fig:ewhale}.  This figure corresponds specifically to the
method of spontaneously  breaking the $U(1)_{\rm L+R}$ symmetry. In
particular, we explicitly show the condensate field, $K$.

Note that the two $SU(5)$ gauge groups are not asymptotically free,
which is a typical problem in realizations of the FN mechanism
\cite{Leurer:1992wg}. In our framework, however, the absence of
asymptotic freedom does not pose a problem since the theory is defined
at the string scale, where new heavy degrees of freedom are integrated in.
The two $SU(5)$ factors are broken into the diagonal $SU(5)$ after
giving a VEV to the Froggatt-Nielsen field $S$. The fields $\bf 10_i$,
$\bf \bar 5_i$, $H_u$ and $H_d$, together with $\bar S$ that becomes
an adjoint, are the matter content of $SU(5)$ GUT. The fields denoted
by $V_i$ are the vector-like fields discussed in section
\ref{sec:frogg-niels-models}.

All other fields are necessary for non-Abelian gauge anomaly
cancellation. The three $X$ fields are connected to a $U(1)$ gauge
group. However, to cancel the anomalies, we can alternatively employ a
single field connected to a $U(3)$ group. All these additional fields
have mass terms in the superpotential and can be integrated out in
pairs: $X_i$ with $\bar X_i$ and $\bar A_{15}$ with the symmetric part
of ${\bf10}_2$.  At the level of the massless spectrum, these fields
may obtain masses through couplings to $S$ or to other moduli which
obtain VEVs and therefore, by definition, are not shown in our
effective quiver.

Reverse geometrically engineering a singular string theory background
with this quiver is a complicated task. A generic construction is
known for a very limited number of cases
\cite{Berenstein:2002ge, Feng:2000mi, Feng:2001xr}.
In fact, we view our model as an effective quiver obtained by Higging
a larger quiver.  One reason for this is that the (symmetric) $\bar
A_{15}$ field cannot be directly obtained together with the
(antisymmetric) $\bf \overline{10}$ field through the orientifold
projection.  Nevertheless, the symmetric $\bar A_{15}$ can originate
from a broken $SU(5)\times SU(5)\to SU(5)$ if a $\bfive\times \bfive$
field is broken down to a $\bar A_{15}$ and a $\overline \ten$, where
the latter is integrated out with another $\ten$.

Thus indeed this theory produces, at low energy, the minimal $SU(5)$
with the correct hierarchy in the up and down sector and with the
predicted neutrino anarchy.

\begin{table}[p]
  \begin{tabular}{l|c}
    Charges &\ Suppression  \\
    \hline \hline
    $(-1,+1)(-1,+1)(-1,+1)$ & 3 \\
    $(0,0)(-1,+1)(-1,+1)$ & 2 \\
    $(-1,0)(0,+1)(-1,+1)$ & 2 \\
    
    $(0,0)(0,0)(-1,+1)$ & 1 \\
    $(0,0)(-1,0)(0,+1)$ & 1 \\
    $(0,+1)(0,-1)(-1,+1)$ & 1 \\
    $(+1,0)(-1,0)(-1,+1)$ & 1 \\
    $(+1,-1)(-1,+1)(-1,+1)$ & 1 \\
    
    \hline
  \end{tabular}
  \caption{Possible $U(1)_{\rm L}\times U(1)_{\rm R}$ charges and the
    resulting suppression factors (namely the power in $\epsilon$) in
    oriented strings.}
  \label{tab:charges-orbifolds}
\end{table}

\begin{table}[p]
    \begin{tabular}{l|c}
      Charges &\ Suppression  \\
      \hline \hline
      $(-2,0)(0,+2)(-1,+1)$ & 3 \\
 
      $(-2,0)(0,+2)(0,0)$ & 2 \\
      $(-2,0)(+1,+1)(-1,+1)$ & 2 \\
      $(-2,0)(0,+1)(0,+1)$ & 2 \\
      $(0,+2)(-1,-1)(-1,+1)$ & 2 \\
      $(0,+2)(-1,0)(-1,0)$ & 2 \\
 
      $(-2,0)(0,+2)(+1,-1)$ & 1 \\
      $(-2,0)(+2,0)(-1,+1)$ & 1 \\
      $(-2,0)(+1,+1)(0,0)$ & 1 \\
      $(-2,0)(+1,0)(0,+1)$ & 1 \\
      $(0,+2)(0,-2)(-1,+1)$ & 1 \\
      $(0,+2)(-1,-1)(0,0)$ & 1 \\
      $(0,+2)(-1,0)(0,-1)$ & 1 \\
      $(+1,+1)(-1,-1)(-1,+1)$ & 1 \\
      $(+1,+1)(-1,0)(-1,0)$ & 1 \\
      $(-1,-1)(0,+1)(0,+1)$ & 1 \\
 
      \hline
    \end{tabular}
    \caption{Possible charges and suppression factors (the power of
      $\epsilon$) with $S(+1,-1)$ that are specific to the unoriented
      case. The sets of table \ref{tab:charges-orbifolds} are also allowed.}
    \label{tab:charges-orientifolds}
 \end{table}

\begin{table}[p]
   \begin{tabular}{l|c}
     Charges &\ Suppression  \\
     \hline \hline
     $(-2,0)(-2,0)(-2,0)$ & 3 \\

     $(-2,0)(-2,0)(0,0)$ & 2 \\
     $(-2,0)(-1,-1)(-1,1)$ & 2 \\
     $(-2,0)(-1,0)(-1,0)$ & 2 \\

     $(-2,0)(-2,0)(+2,0)$ & 1 \\
     $(-2,0)(0,-2)(0,+2)$ & 1 \\
     $(-2,0)(-1,-1)(+1,+1)$ & 1 \\
     $(-2,0)(0,0)(0,0)$ & 1 \\
     $(-2,0)(-1,0)(+1,0)$ & 1 \\
     $(-2,0)(0,-1)(0,+1)$ & 1 \\
     $(-2,0)(-1,+1)(+1,-1)$ & 1 \\
     $(0,-2)(-1,+1)(-1,+1)$ & 1 \\
     $(0,2)(-1,-1)(-1,-1)$ & 1 \\
     $(-1,-1)(0,0)(-1,+1)$ & 1 \\
     $(-1,-1)(-1,0)(0,+1)$ & 1 \\
     $(0,0)(-1,0)(-1,0)$ & 1 \\
     $(-1,0)(0,-1)(-1,+1)$ & 1 \\
     \hline
   \end{tabular}
   \caption{Possible charges and suppression factors (the power of
     $\epsilon$) in the unoriented case with $S(+2,0)$. The $S(-2,0)$
     case is obtained by multiplying the $U(1)_{\rm L}$-charge by $-1$.}
   \label{tab:charges-orientifolds3}
\end{table}

\begin{table}[p]
  \begin{center}
    \begin{tabular}{l|c|c|c}
      $SU(5)$ & Model & $U(1)_L\times U(1)_R$ & $U(1)_{L-R}$\\
      \hline\hline
      $10_i$ $(i=1,2,3)$ & $T_1$ & $(2,0)(2,0)(2,0)$ & $(+2,+2,+2)$ \\
      & $T_2$ & $(2,0)(2,0)(0,2)$ & $(+2,+2,-2)$\\
      & $T_3$ & $(2,0)(2,0)(1,1)$ & $(+2,+2,0)$\\
      & $T_4$ & $(2,0)(0,2)(0,2)$ & $(+2,-2,-2)$\\
      & $T_5$ & $(2,0)(0,2)(1,1)$ & $(+2,-2,0)$\\
      & $T_6$ & $(2,0)(1,1)(1,1)$ & $(+2,0,0)$\\
      & $T_7$ & $(0,2)(0,2)(0,2)$ & $(-2,-2,-2)$\\
      & $T_8$ & $(0,2)(0,2)(1,1)$ & $(-2,-2,0)$\\
      & $T_9$ & $(0,2)(1,1)(1,1)$ & $(-2,0,0)$\\
      & $T_{10}$ & $(1,1)(1,1)(1,1)$ & $(0,0,0)$\\
      &&&\\
      $\bar 5_i$ $(i=1,2,3)$ & $F_1$ & $(-1,0)(-1,0)(-1,0)$ & $(-1,-1,-1)$\\
      & $F_2$ & $(-1,0)(-1,0)(0,-1)$ & $(-1,-1,+1)$\\
      & $F_3$ & $(-1,0)(0,-1)(0,-1)$ & $(-1,+1,+1)$\\
      & $F_4$ & $(0,-1)(0,-1)(0,-1)$ & $(+1,+1,+1)$\\
      &&&\\
      $H_d(\bar 5)$, $h_u(5)$ & $D_1$ & $(-1,0),(+1,0)$ & $(-1,+1)$\\
      & $D_2$ & $(-1,0),(0,+1)$ & $(-1,-1)$\\
      & $D_3$ & $(0,-1),(+1,0)$ & $(+1,+1)$\\
      & $D_4$ & $(0,-1),(0,+1)$ & $(+1,-1)$\\
      &&&\\
      $H_u(5)$, $h_d(\bar5)$& $U_1$ & $(-4,0),(+4,0)$ & $(-4,+4)$\\
      & $U_2$ & $(-4,0),(0,+4)$ & $(-4,-4)$\\
      & $U_3$ & $(0,-4),(+4,0)$ & $(+4,+4)$\\
      & $U_4$ & $(0,-4),(0,+4)$ & $(+4,-4)$\\
      &&&\\
      $S(1)$ & & $(+1,-1)$ & $+2$\\
      \hline
    \end{tabular}
  \end{center}
  \caption{All possible charge assignments for the model with
      additional fields and an unbroken $U(1)_{L+R}$ symmetry} 
  \label{tab:2HM-composite}
\end{table}

\begin{table}[p]
  \begin{center}
    \begin{tabular}{l|c|c|c}
      $SU(5)$ &\ Model &\ $U(1)_L\times U(1)_R$ &\ $U(1)_{L-R}$\\
      \hline\hline
      $10_i$ & $T_2$ & $(2,0)(2,0)(0,2)$ & $(+2,+2,-2)$\\
             & $T_4$ & $(2,0)(0,2)(0,2)$ & $(+2,-2,-2)$\\
             & $T_5$ & $(2,0)(0,2)(1,1)$ & $(+2,-2,0)$\\
             & $T_6$ & $(2,0)(1,1)(1,1)$ & $(+2,0,0)$\\
             & $T_9$ & $(0,2)(1,1)(1,1)$ & $(-2,0,0)$\\
      &&&\\
      $\bar 5_i$ & $F_1$ & $(-1,0)(-1,0)(-1,0)$ & $(-1,-1,-1)$\\
                 & $F_2$ & $(-1,0)(-1,0)(0,-1)$ & $(-1,-1,+1)$\\
                 & $F_3$ & $(-1,0)(0,-1)(0,-1)$ & $(-1,+1,+1)$\\
                 & $F_4$ & $(0,-1)(0,-1)(0,-1)$ & $(+1,+1,+1)$\\
      &&&\\
      $H_d(\bar 5)$ & $D_1$ & $(-1,0)$ & $-1$\\
                    & $D_2$ & $(0,-1)$ & $+1$\\
      &&&\\
      $H_u(5)$ & $U_1$ & $(1,0)$ & $+1$\\
               & $U_2$ & $(0,1)$ & $-1$\\
      &&&\\
      $S(1)$ & &$(+1,-1)$ & $+2$\\
      \hline
    \end{tabular}
  \end{center}
  \caption{All possible charge assignments for the model with
      instanton breaking of the $U(1)_{L+R}$}
  \label{tab:instbreak}
\end{table}

\begin{table}[p]
  \begin{center}
    \begin{tabular}{l|c|c|c}
      $SU(5)$ & Model & $U(1)_L\times U(1)_R$ & $U(1)_{L-R}$\\
      \hline\hline
      $10_i$ & $T_1$ & $(2,0)(2,0)(2,0)$ & $(+2,+2,+2)$ \\
      & $T_2$ & $(2,0)(2,0)(0,2)$ & $(+2,+2,-2)$\\
      & $T_3$ & $(2,0)(2,0)(1,1)$ & $(+2,+2,0)$\\
      & $T_4$ & $(2,0)(0,2)(0,2)$ & $(+2,-2,-2)$\\
      & $T_5$ & $(2,0)(0,2)(1,1)$ & $(+2,-2,0)$\\
      & $T_6$ & $(2,0)(1,1)(1,1)$ & $(+2,0,0)$\\
      & $T_7$ & $(0,2)(0,2)(0,2)$ & $(-2,-2,-2)$\\
      & $T_8$ & $(0,2)(0,2)(1,1)$ & $(-2,-2,0)$\\
      & $T_9$ & $(0,2)(1,1)(1,1)$ & $(-2,0,0)$\\
      & $T_{10}$ & $(1,1)(1,1)(1,1)$ & $(0,0,0)$\\
      &&&\\
      $\bar 5_i$ & $F_1$ & $(-1,0)(-1,0)(-1,0)$ & $(-1,-1,-1)$\\
      & $F_2$ & $(-1,0)(-1,0)(0,-1)$ & $(-1,-1,+1)$\\
      & $F_3$ & $(-1,0)(0,-1)(0,-1)$ & $(-1,+1,+1)$\\
      & $F_4$ & $(0,-1)(0,-1)(0,-1)$ & $(+1,+1,+1)$\\
      &&&\\
      $H_d(\bar 5)$ & $D_1$ & $(-1,0)$ & $-1$\\
      & $D_2$ & $(0,-1)$ & $+1$\\
      &&&\\
      $H_u(5)$ & $U_1$ & $(+1,0)$ & $+1$\\
      & $U_2$ & $(0,+1)$ & $-1$\\
      &&&\\
      $K(1),\bar K(1)$ & $K_1$ & $(+5,0),(-5,0)$ & $(+5,-5)$\\
      & $K_2$ & $(0,+5),(0,-5)$ & $(-5,+5)$\\
      &&&\\
      $S(1)$ & &$(+1,-1)$ & $+2$\\
      \hline
    \end{tabular}
  \end{center}
  \caption{All possible charge assignments for the model with
       spontaneous breaking of the $U(1)_{L+R}$ symmetry}
  \label{tab:spontanbreak}
\end{table}


\end{document}